\documentclass[onecolumn]{article}

\usepackage{indentfirst}
\usepackage{xcolor}
\usepackage{graphicx}
\usepackage{url}

\title{Decentralized Verifiable Mail-in Ballot Counting for Postal Voting}
\author{Peichen Xie$^1$ \and Zihan Zheng$^2$ \and Xian Zhang$^1$ \and Shuo Chen$^1$}
\date{$^1$Microsoft Research\\
$^2$University of Science and Technology of China}

\begin{document}
\maketitle

\begin{abstract}
As computer vision is prevalently used for mail-in ballot processing and counting, it becomes a point of centralized trust in postal voting. We propose DVote, a prototype system of postal voting that provides decentralized trust in computer vision. With blockchain and layer-2 technologies, DVote decentralizes the computation and model training of computer vision to a group of scrutineers that hold the AnyTrust assumption, i.e., at least one member is honest. Consequently, the computational integrity is anchored to the trustworthiness of a large public blockchain such as Ethereum.
\end{abstract}
\section{Introduction}

Voting is by nature a process based on decentralized trust because no single entity can be fully trusted by all voters and parties. 
As \textit{postal voting} is more prevalent, the large number of mail-in ballots are processed and counted by machines in the election office. As shown in public videos~\cite{youtube2,youtube1}, the machines use computer vision to verify handwritten signatures and read votes from the ballots. Traditionally, the procedure must be undertaken and watched by humans from opposing parties, embodying decentralized trust. However, today, computer vision largely takes over humans in this procedure and becomes the point of centralized trust. The general public is asked to trust the outcomes produced by the machines. This is highly debatable, regardless of whether there is any actual fraud or cyber attack in the procedure. In light of the serious trust crisis after the 2016 and 2020 US general elections, it is clear that the public needs confidence in the procedure.

In this paper, we present DVote, a prototype system of postal voting with a core function of decentralized verifiable mail-in ballot counting. The core idea is to use blockchain and layer-2 technologies to decentralize the computation of the computer vision algorithms in ballot counting. As a result, the trustworthiness of the computation is similar to the trustworthiness of a large public blockchain (such as Ethereum).

The most important assumption of DVote is AnyTrust~\cite{kalodner_arbitrum_2018}, which essentially means that the scrutineers (i.e., the decentralized nodes performing the computation) are not fully corrupted. 
This is a natural assumption in voting because the scrutineers consist of multiple (even opposing) parties and entities.

By leveraging the opposing forces, DVote guarantees the computation in ballot counting, especially the signature verification and vote reading of mail-in ballots, is genuine. On the other hand, DVote also guarantees the computer vision model used for the computation is audited beforehand, by providing proof of provenance that commits how the model is trained. The underlying technology of both components is called Agatha~\cite{zheng_agatha_2021}, which enables decentralized verifiable deep learning computation and bridges the gap between AnyTrust and the trustworthiness of blockchains.

The techniques for decentralized verification are non-trivial. First, rigorous verification requires the computation to be deterministic, so we take many efforts to ensure the (bit-wise) reproducibility of deep learning computation. Second, the computational performance must remain reasonably high in order to process the numerous ballots, so we utilize fraud proof instead of cryptography methods for decentralized verification. At last, with these challenges tackled, the computation will be done with negligible slowdown while rational scrutineers will not cheat because dishonest results will always be disproven by an honest scrutineer (thanks to AnyTrust).

In the following part of this paper, we first describe the overview of the DVote prototype and then detail the two important components of DVote, i.e., decentralized verifiable ballot counting (DVBC) and decentralized algorithm audit (DAA). We also note that there are numerous vulnerabilities besides computational integrity in postal voting \cite{heinl_comparative_2021,killer_swiss_2019}. End-to-end verifiability is not the goal of this paper, so we leave those vulnerabilities for further study.

\section{Overview of DVote}

In this section, we present the overview of DVote, our prototype system for postal voting. DVote includes five stages, which constitute a simplified process of postal voting. DVBC and DAA, the most important components of DVote, are integrated into the ballot counting stage and the setup stage respectively. In addition to our description, readers may refer to \cite{heinl_comparative_2021,killer_swiss_2019} for the detailed real-world postal voting systems.

\subsection{Voting process}

\begin{figure}[h]
\centering
\includegraphics[width=\columnwidth]{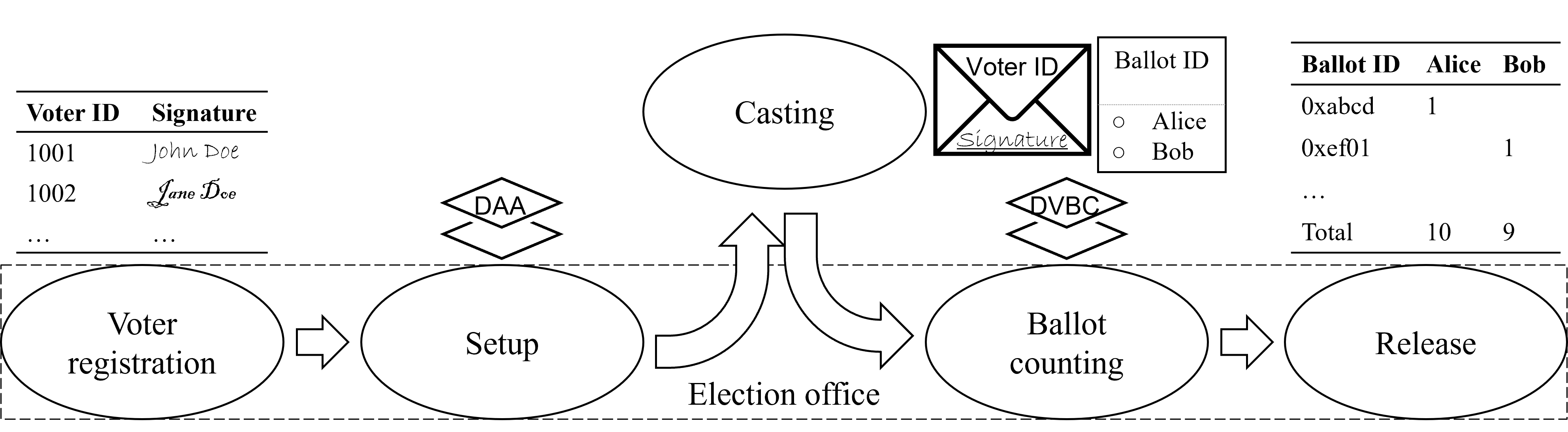}
\caption{Five-stage process of postal voting}
\label{fig:voting_process}
\end{figure}

Figure \ref{fig:voting_process} shows the general five-stage process of DVote. 
The first stage, voter registration, is a pre-voting stage, where eligible voters register necessary personal information. Here we highlight the IDs and signatures as they will be critical in ballot counting.

The second stage includes the setup of the election office, production of mail-in ballots (including ballot papers and voting envelopes\footnote{There may be extra ballot envelopes, signature card, etc.}), and the delivery of the ballots. Each envelope is printed with a voter ID, and each ballot paper is printed with a random ballot ID\footnote{The ballot ID is optional as traceable ballots are prohibited in some places. Removing ballot IDs can make the ballots untraceable.}. The voter ID and the ballot ID should be unlinkable.

In the third stage, the voter receives the mail-in ballot. He/she should mark his/her vote on the ballot paper, fold the ballot paper, put it into the envelope, and sign his/her name on the envelope. Then, the ballot is posted to the election office.

In the fourth stage, the election office receives the mail-in ballots and counts the ballots with the machines. We will detail this stage in Section~\ref{sec:vote_counting}.

After ballot counting, the result of voting will be published. At this last stage, voters can review the correctness of the result and check whether their individual ballot is correctly recorded. We do not focus on ballot secrecy for the moment, because this is orthogonal to this paper. Ballot secrecy is an arguing issue in postal voting, and it could be mitigated by removing ballot IDs or using cryptography \cite{benaloh_verifiable_2013}.

\subsection{Ballot counting}
\label{sec:vote_counting}

\begin{figure}[h]
\centering
\includegraphics[width=0.7\columnwidth]{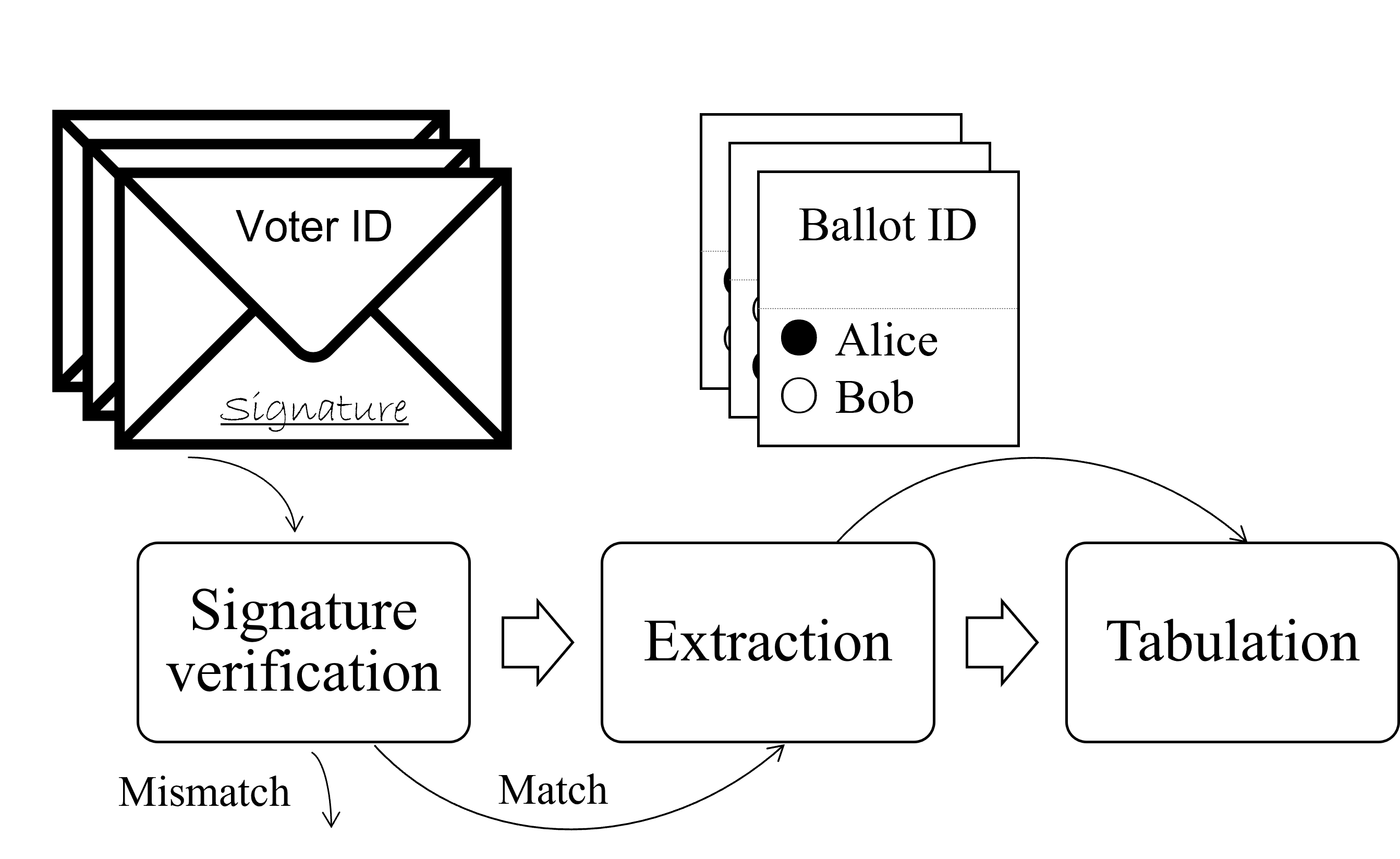}
\caption{Three-step process of ballot counting}
\label{fig:vote_counting}
\end{figure}

Within the five stages, the ballot counting stage is the focus of this paper. Figure \ref{fig:vote_counting} shows the three-step process of mail-in ballot counting. 

The mail-in ballots are put into a sorting machine at the first step. For each ballot, the voter ID's bar code and the handwritten signature are scanned by the machine, and the signature is compared with the registered signature of this voter ID, by a specific handwritten signature verification algorithm. This algorithm is audited in the setup stage with DAA, as we will describe in Section~\ref{sec:provenance}, and is executed in a decentralized verifiable manner with DVBC, as we will describe in Section~\ref{sec:design}.

On, for example, the election day, the verified ballots are sent to a machine to open the envelopes and extract the ballot papers into ballot boxes. 
After this step, the voting envelopes and the ballot papers are unlinked. No one knows the mapping between the voter IDs and the ballot IDs except the voters themselves.

Then, the ballot papers are sent to a vote tabulator for optical scanning and tabulation. Only a small number of unidentifiable ballots may need manual double-check. Otherwise, the votes are automatically read and counted. Similar to handwritten signature verification, the algorithm used in this step is also secured by the techniques of DVBC and DAA.
\section{Decentralized Verifiable Ballot Counting}
\label{sec:design}

In this section, we present DVBC, a component that enables decentralized verifiable computing for mail-in ballot counting.

\subsection{Target of DVBC}

In the ballot counting process described in the last section, computer vision is critical for handwritten signature verification and vote tabulation. They both can be divided into two sub-steps:  scanning and image processing. Image scanning is assumed uncontroversial because all the paper records are stored and auditable. However, the image processing today is done with non-transparent algorithm and unverifiable computation.

How can people verify the integrity of the image processing? The computation can be outsourced to a fully trusted and verifiable server if there is one. However, no single entity can be trusted by everyone in voting. Thus, in this section, we aim to make the computation verifiable in a decentralized manner. 

We assume people have agreed on what computer vision algorithm should be used in this section. In Section \ref{sec:provenance}, we will further discuss the provenance of the algorithm. 

\subsection{Design of DVBC}
\label{sec:verification}

We firstly employ a large public blockchain with general computation capability for the computation in ballot counting. We take Ethereum for example. The (quasi) Turing completeness of Ethereum virtual machine~\cite{gavin_wood_ethereum_2021} enables arbitrary computation to be executed and verified on \textit{all} blockchain nodes, provided that the amount of computation is under the cost (i.e., the gas in Ethereum) limit. As a result, the nodes as a decentralized community can reach an agreement on whether the computation is correct, although no member in the community is trusted by everyone. The trust assumption for community consensus is about the whole community, as opposed to an individual. For example, the assumption can be that at least half of the community is honest.

\begin{figure}[h]
\centering
\includegraphics[width=0.8\columnwidth]{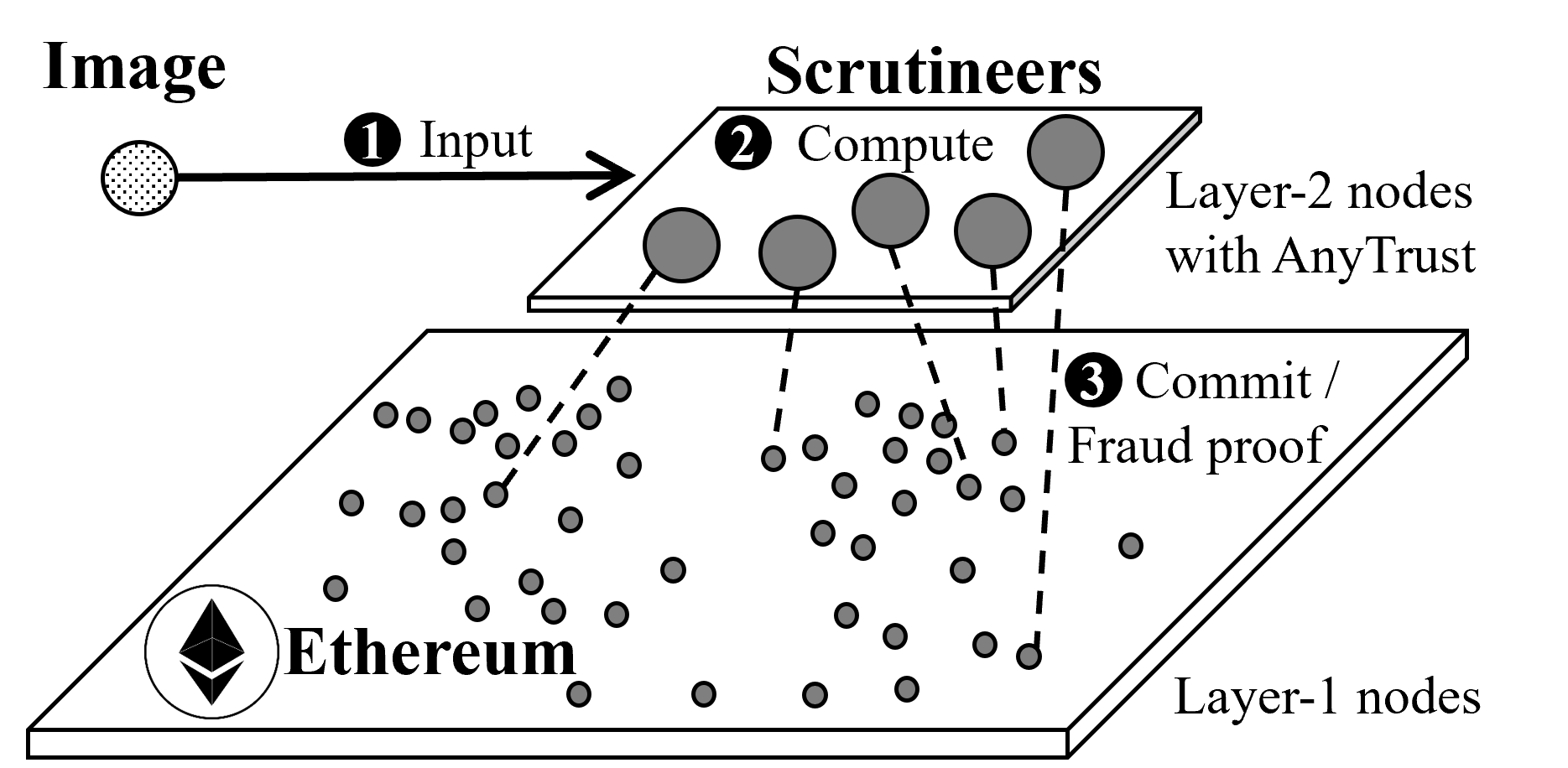}
\caption{Verifiable computing with decentralized trust}
\label{fig:verification}
\end{figure}

Secondly, we employ layer-2 techniques~\cite{gudgeon_sok_2020} to handle expensive computer vision algorithms, such as the deep learning models used in handwritten signature verification for feature extraction. Imagine a VGG16 model~\cite{simonyan_very_2015} containing 30 billion floating-point operations runs on Ethereum. Even with the minimum gas consumption for every operation, we estimate that processing a single image will consume 90 billion gas, which exceeds the block gas limit by 3000 times. Therefore, we use Agatha~\cite{zheng_agatha_2021} to offload the computation to a small group of scrutineers (i.e., layer-2 nodes as shown in Figure~\ref{fig:verification}), assuming at least one scrutineer is honest (i.e., AnyTrust), so that Ethereum only needs to do simple acknowledgment or arbitration. Specifically, if the scrutineers' results are consistent, Ethereum accepts the result without re-computation; otherwise, Ethereum arbitrates between inconsistent scrutineers by an $O(\log n)$ fraud proof protocol~\cite{kalodner_arbitrum_2018,zheng_agatha_2021}.

AnyTrust is a reasonable assumption in voting. It is natural to choose the scrutineers to represent the various parties of the candidates and different entities such as the government, the congress, the court, the media, etc. All these entities may want to cheat in order to benefit their favored candidates. However, the AnyTrust assumption still holds, because \textit{for each ballot}, at least one scrutineer is motivated to be truthful. In other words, it is logical to believe that no scrutineer is motivated to falsely approve an invalid ballot of its opponent candidate or falsely disapprove a valid ballot of its favored candidate.
 
A potential serious misunderstanding of our design is that Ethereum delegates the group of layer-2 nodes to reach a consensus for them. Suppose 7 out of 10 nodes are dishonest and they claim ``$1+1=3$". If Ethereum delegated the 10 nodes to reach a consensus, ``$1+1=3$" would become 10 nodes’ consensus in the proof of work mechanism (i.e., the simple majority rule). However, this is not what happens in DVBC. There is no concept called ``the scrutineers' consensus" when their results are inconsistent. Under the AnyTrust assumption, an honest node will succeed in disproving ``$1+1=3$" in front of Ethereum, and the claim is thus rejected. On the other hand, \textit{assuming AnyTrust}, if all nodes claim ``$\textrm{VGG16}(x)=y$", this claim is as trustworthy as ``$\textrm{VGG16}(x)=y$" is computed by Ethereum, but the computational cost for the former is negligible compared to the latter.

\section{Decentralized Algorithm Audit}
\label{sec:provenance}

In this section, we aim to design a mechanism to audit the computer vision algorithm used in mail-in ballot counting. As we noted, the image processing in mail-in ballot counting today is done in a centralized manner, with a non-transparent algorithm and unverifiable computation. Although we have guaranteed \textit{computational} integrity in Section~\ref{sec:design}, the algorithm itself may have vulnerabilities (e.g., back doors~\cite{shafahi_poison_2018}).

To address this issue, we present DAA, a component added into the setup stage in DVote. Traditionally, programs can be audited via code review. However, deep learning models cannot, as only their parameters and architecture are visible. People cannot tell how the parameters are trained, whether the training data are poisoned, or whether they are injected with back doors during training. Thus, the model \textit{provenance} should also be reviewed in a decentralized manner.

As if in DVBC, we leverage a decentralized group of scrutineers to audit the model provenance, by verifying the computation of model training. We express the training as
\begin{equation}
\label{eq:training}
    M=f(...f(f(M_0,D_1),D_2)...D_n)
\end{equation}
where $M$ is the final model, $M_0$ is the initial model, $n$ is the number of iterations, $f$ is the training algorithm (e.g., gradient descent), and $D_i$ is called a mini-batch, representing the training data used in the $i$-th iteration. Given $M_0$, $\{D_i\}$ and $f$, the scrutineers verifies the computation~(\ref{eq:training}) by the method detailed in Section~\ref{sec:verification}.

Specifically, both DVBC and DAA use Agatha~\cite{zheng_agatha_2021} for decentralized verifiable deep learning computation: DVBC for model ``inference" but DAA for model training. Since the original Agatha only supports deterministic computational graphs, to verify model training with Agatha, computation~(\ref{eq:training}) must be deterministic even if it runs on GPUs. We address this issue by  integrating reproducible deep learning techniques~\cite{repro} into Agatha's ``sequencing" technique and leveraging the data parallelism in matrix multiplications and deep learning. 

As a result of verifiable training, people can have a consensus that $M$ is correctly trained from $M_0$, $\{D_i\}$ and $f$. Anyone can review this provenance by querying a scrutineer he/she trusts. Thanks to Agatha, the cost of DAA is negligible compared with other verifiable computing methods.

Note that the intrinsic weakness of deep learning models is out of this paper's scope. It should be enhanced in other ways. For example, people are using deep learning models today for handwritten signature verification although they cannot achieve $100\%$ accuracy because a manual double-check can mitigate the problem. One common concern about making the models non-secret is adversarial examples~\cite{szegedy_intriguing_2014}. This vulnerability can be mitigated by ensemble methods. 
\section{Conclusion}

Computation in postal voting can be made verifiable with decentralized trust. To this end, we propose the DVote prototype with blockchain technologies and the AnyTrust assumption, ensuring that the computation in mail-in ballot counting is as trustworthy as a large public blockchain. We hope this study can inspire the decentralized computing paradigm to target challenging problems in society.

\bibliographystyle{plain}
\bibliography{references}

\end{document}